\documentstyle[12pt]{article}
\begin{document}

\title{Gauge and parametrization dependence in higher derivative quantum gravity}

\author{~Kirill~A.~Kazakov\thanks{E-mail: $kirill@theor.phys.msu.su$} ~and
~Petr~I.~Pronin\thanks{E-mail: $petr@theor.phys.msu.su$}}

\maketitle

\begin{center}
{\em Moscow State University, Physics Faculty,\\
Department of Theoretical Physics.\\
$117234$, Moscow, Russian Federation}
\end{center}

\begin{abstract}

The structure of counterterms in higher derivative quantum gravity
is reexamined. Nontrivial dependence of charges on the gauge and parametrization
is established. Explicit calculations of two-loop contributions are
carried out with the help of the generalized renormgroup method
demonstrating consistency of the results obtained.

\end{abstract}

\noindent

\vspace{0,5cm}
{\large{\bf 1.Introduction}}
\vspace{0,5cm}

As well known, not all of the problems of the quantum field theory
are exhausted by the construction of S-matrix.
Investigation of evolution of the Universe, behavior of quarks in
quantum chromodynamics etc. require the introduction of more general
object $-$ the so called effective action. Besides that, the program
of renormalization of the S-matrix itself has not yet been carried out
in terms of the S-matrix alone.
Renormalization of the Green functions, therefore, is the central point
of the whole theory. Given these functions one can obtain the S-matrix
elements with the help of the reduction formulas.
In this respect those properties of the generating functionals which
remain valid after the transition to the S-matrix is made are of special
importance.

We mean first of all the properties of the so called
"essential" coupling constants in the sense of S.Weinberg \cite{wein}.
They are defined as those independent from any redefinition of the fields.
In the context of the quantum theory one can say that the renormalization
of "essential" charges is independent from renormalizations of the fields.
Separation of quantities into "essential" and "inessential" ones is
convenient and we use it below.

In this paper we shall consider the problem of gauge and parametrization
dependence of the effective action of $R^2$-gravity.

There are two general and powerful methods of investigation of gauge
dependence in quantum field theory. The first of them \cite{lavtut} uses the
Batalin-Vilkovisky formalism \cite{batvil1,batvil2,batvil3} and is based on the fact that
any change of gauge condition can be presented as a (local) canonical (in the
sense of "antibrackets" \cite{batvil2}) transformation of the effective action.
This canonical transformation induces corresponding renormalized
canonical transformation of the renormalized effective action.
This leads to the following result: the renormalization boils down to the
redefinition of the coupling constants (which are the coefficients of
independent gauge invariant structures entering the Lagrangian)
and some canonical transformation of the fields and sources of BRST-transformations.
The second approach \cite{kluberg1, kluberg2, stelle}
consists in the introduction of some additional anticommuting source
to the effective action in such a way that the Slavnov identities
for the corresponding generating functional of proper vertices connect
its derivatives with respect to the gauge fixing parameter and to the mean
fields (and sources of BRST-transformations).

The second method was used in \cite{stelle} to prove the gauge independence
of the gauge invariant divergent parts of the effective action up to
the terms proportional to the classical equations of motion of the gravitational
field. Together with the general result of the first approach
mentioned above this would imply some far-reaching consequences
concerning renormalization of the fields. For example, one could conclude
that the canonical transformation corresponding to a change
of the gauge condition should not be renormalized. Unfortunately,
this is not the case.
We will show in this paper that the aforesaid result of \cite{stelle}
holds at the one-loop level only.
Introduction of the additional source mentioned above requires also
introducing of some additional terms needed to compensate divergences
which arise because of the presence of the new source. As a result
corresponding Slavnov identities impose only some constraints on the divergent
structures from which a nontrivial dependence\footnote{I.e. dependence which
can not be presented as proportional to the equations of motion.}
on the gauge follows already at the two-loop level.

Our paper is organized as follows. In sec.2 we determine possible
divergent structures which are originated due to the presence of the new source
and obtain correct Slavnov identities. In sec.3 we
calculate explicitly the divergent part of the effective action
at the one loop level in arbitrary (linear) gauge and the special class
of parametrizations. In sec.4 we calculate the divergent
as $\frac{1}{\varepsilon^2}$ ($\varepsilon$ being the dimensional
regulator\footnote{We set $2\varepsilon = d - 4$, $d$ being the dimensionality of the space-time.})
part at the two-loop level with the help of the generalized
renormgroup method and show that the results obtained in secs.3,4 satisfy
the relations derived in sec.2.

We use highly condensed notations of DeWitt throughout this paper.
Also left derivatives with respect to anticommuting variables are used.
The dimensional regularization of all divergent quantities is supposed.

\vspace{1cm}

{\large{\bf 2. Generating functionals and Slavnov identities}}

\vspace{0,5cm}

{\bf a. Action, gauge fixing and parametrization}

\vspace{0,5cm}

Let us consider higher derivative quantum gravity described
by an action which includes the minimal set of terms added
to the usual ones of Einstein to ensure the power counting
renormalizability of the theory\footnote{Our notations are
$R_{\mu\nu} \equiv R^{\alpha}_{\mu\alpha\nu} =
\partial_{\alpha}\Gamma^{\alpha}_{\mu\nu} - ...,
R \equiv R_{\mu\nu} g^{\mu\nu}, g_{\mu\nu} = sign(+,-,-,-).$}

\begin{eqnarray}\label{eqofmot}&&
S_{0} = \int\,d^4 x\,\sqrt{-g}(\alpha_1 R^2 + \beta R_{\mu\nu} R^{\mu\nu} - \frac{1}{k^2} (R - 2 \Lambda)),
\end{eqnarray}

where $\alpha_1$ and $\beta$ are arbitrary constants satisfying only
$\beta \ne 0$, $3\alpha_1 + \beta \ne 0$  which imply that the graviton propagator
behaves like $p^{-4}$ for large momenta (see \cite{stelle});
$k$ is the gravitational constant and $\Lambda$ is the cosmological term.

The corresponding equations of motion are

\begin{eqnarray}\label{eqmot}&&
\frac{1}{2}\alpha_1 R^2 g^{\alpha\beta} + \frac{1}{2}\beta R^{\mu\nu} R_{\mu\nu}g^{\alpha\beta} - 2\alpha_1 R R^{\alpha\beta}
\nonumber\\&&
- 2\beta R_{\mu\nu}R^{\mu\alpha\nu\beta} - \left(2\alpha_1 + \frac{1}{2}\beta\right)\Box R g^{\alpha\beta} - \beta\Box R^{\alpha\beta}
\nonumber\\&&
+ (2\alpha_1 + \beta)R^{;\alpha\beta} - \frac{1}{2 k^2}R g^{\alpha\beta} + \frac{1}{k^2}\Lambda g^{\alpha\beta} + \frac{1}{k^2}R^{\alpha\beta}  = 0.
\end{eqnarray}

Renormalizability of this theory was proved in \cite{stelle} in the case
of the so called unweighted (or weighted with a functional containing
fourth or higher derivatives) harmonic gauge condition.
The proof in more general case boils down to the proof of the
so called locality hypotheses. In \cite{voronovtutin} its validity
was shown most generally.

For our purposes it is sufficient to consider
the harmonic gauge\footnote{We use the flat-space metric tensor
$\eta^{\mu\nu} = diag(+1,-1,-1,-1)$ to raise Lorentz indices.} following Stelle \cite{stelle}

\begin{eqnarray}\label{gauge}&&
F_{\mu} \equiv F_{\mu}^{\alpha\beta}h_{\alpha\beta} \equiv
\partial^{\nu}h_{\mu\nu} = 0,
\end{eqnarray}

where $h_{\mu\nu}$ denotes some set of dynamical variables describing
the gravitational field. We recall that in the theory of gravity
a natural ambiguity in the choice of such a set exists because the
generators $D_{\mu\nu}^{\alpha}$ of gauge transformations of variables constructed from
the metric $g_{\mu\nu}$ (or $g^{\mu\nu}$) and its determinant $g = detg_{\mu\nu}$
in any combination have simple form linear in fields and their derivatives.
For general constructions of this section it doesn't matter what choice
we make. We only note that the gauge in the form of (\ref{gauge})
will always correspond to the set of dynamical variables\footnote{They will be
referred to as {\it standard variables}.} which
enter the so called reduced expression of the metric expansion (see sec.3a, eq.(\ref{standard})).
Thus the BRST-transformations of the Faddeev-Popov effective action with the gauge
fixing term being $-\frac{1}{2\Delta}F^{\alpha}\Box F_{\alpha}$,
expressed in terms of these standard variables are

\begin{eqnarray}\label{brs}&&
\delta h_{\mu\nu} = D_{\mu\nu}^{\alpha} C_{\alpha}\lambda,
\nonumber\\&&
\delta C_{\alpha} = - \partial^{\beta}C_{\alpha}C_{\beta}\lambda,
\nonumber\\&&
\delta \bar{C}^{\tau} = - \Delta^{-1}\Box F^{\tau}\lambda,
\end{eqnarray}

where $\lambda$ is an anticommuting constant parameter.

\vspace{0,5cm}

{\bf b. Green functions}

\vspace{0,5cm}

We write the generating functional of Green functions in the extended form
of Zinn-Justin \cite{zinnjustin} modified by Kluberg-Stern
and Zuber \cite{kluberg1,kluberg2}

\begin{eqnarray}\label{gener}&&
Z[T^{\mu\nu},\bar{\beta}^{\sigma},\beta_{\tau},K^{\mu\nu},L^{\sigma}] =
\nonumber\\&&
\hspace{-0,5cm}
{\displaystyle\int}dh_{\mu\nu}dC_{\sigma}d\bar{C}^{\tau}exp\{i (\tilde{\Sigma}(h_{\mu\nu},C_{\sigma},\bar{C}^{\tau},K^{\mu\nu},L^{\sigma})
\nonumber\\&&
+ Y F_{\sigma}\bar{C}^{\sigma} + \bar{\beta}^{\sigma}C_{\sigma} + \bar{C}^{\tau}\beta_{\tau} + T^{\mu\nu}h_{\mu\nu})\}
\end{eqnarray}

where $$\tilde{\Sigma} = S_{0} - \frac{1}{2\Delta}F^{\alpha}\Box F_{\alpha}
+ \bar{C}^{\tau}F_{\tau}^{\mu\nu}D_{\mu\nu}^{\alpha}C_{\alpha}
+ K^{\mu\nu}D_{\mu\nu}^{\alpha}C_{\alpha}
+ L^{\sigma}\partial^{\beta}C_{\sigma}C_{\beta};$$

$K^{\mu\nu}(x)$ (anticommuting), $L^{\sigma}(x)$ (commuting) are
the BRST-transformation sources and Y is a constant anticommuting parameter.

Let us first consider the structure of divergencies which correspond to the
extra source Y. Power counting gives for the degree of divergence D of an
arbitrary diagram
\begin{equation}\label{degree}
D = 4 - 2 n_2 - 2 n_K - n_L - 2 n _Y - E_C - 2 E_{\bar{C}},
\end{equation}

where $n_2$ = number of graviton
vertices with two derivatives, $n_{K,L,Y}$ = numbers of K,L,Y vertices
respectively, $E_C$ and $E_{\bar{C}}$ = numbers of external ghost and antighost lines.

Also from the expression (\ref{gener}) we see that one can ascribe the following
ghost numbers $N_g$ to all the fields and sources:

\begin{eqnarray}\label{gnumbers}&&
N_g[h] = 0, N_g[C] = + 1, N_g[\bar{C}] = - 1,
\nonumber\\&&
N_g[K] = - 1, N_g[L] = - 2, N_g[Y] = + 1.
\end{eqnarray}

Now from (\ref{degree}) and (\ref{gnumbers}) one can see that there are
three types of divergent structures involving Y-vertex: Y K, Y $\bar{C}$ and Y L C,
each of which may have arbitrary number of external graviton lines.
As far as we have adopted the standard covariant approach thus only Lorentz
covariant quantities may appear and therefore we have for the general form
of the above structures

\begin{eqnarray}\label{yk}&&
Y K^{\mu\nu} P_{\mu\nu},
\end{eqnarray}

\begin{eqnarray}\label{yc}&&
Y \bar{C}^{\nu}\partial^{\mu} Q_{\mu\nu}
\end{eqnarray}

and

\begin{eqnarray}\label{ylc}&&
Y L^{\sigma}C^{\tau}M_{\sigma\tau},
\end{eqnarray}

where $P$, $Q$ and $M$ are some Lorentz-covariant tensors depending on $h_{\mu\nu}$ alone.

Thus to renormalize the Green functions we must introduce corresponding
counterterms and consider the new generating functional

\begin{eqnarray}\label{genernew}&&
Z[T^{\mu\nu},\bar{\beta}^{\sigma},\beta_{\tau},K^{\mu\nu},L^{\sigma}] =
\nonumber\\&&
{\displaystyle\int}dh_{\mu\nu}dC_{\sigma}d\bar{C}^{\tau}
exp\{i (\tilde{\Sigma}(h_{\mu\nu},C_{\sigma},\bar{C}^{\tau},K^{\mu\nu},L^{\sigma})
\nonumber\\&&
+ Y K^{\mu\nu}P_{\mu\nu}
+ Y \bar{C}^{\nu}\partial^{\mu} Q_{\mu\nu}
\nonumber\\&&
+ Y L^{\sigma}C^{\tau}M_{\sigma\tau} + \bar{\beta}^{\sigma}C_{\sigma}
+ \bar{C}^{\tau}\beta_{\tau} + T^{\mu\nu}h_{\mu\nu})\}
\end{eqnarray}

instead of (\ref{gener}).

\vspace{1cm}

{\bf c. Slavnov identities}

\vspace{1cm}

Let us proceed to successive renormalization of Green functions
corresponding to (\ref{genernew}).
We will first consider the case when $g^{*}_{\mu\nu} = g_{\mu\nu}$ are
chosen as a parametrization of the gravitational field. Then the
general result will be clear.

To ensure renormalizability we work with a generating functional
(\ref{genernew}) from the very beginning. We will see below that Slavnov
identities determine the structure of the polynomials P and Q completely.
They turn out to be

\begin{eqnarray}\label{cont1}&&
P_{\mu\nu} = a (\eta_{\mu\nu} + h_{\mu\nu}),
\end{eqnarray}

\begin{eqnarray}\label{cont2}&&
Q_{\mu\nu} = a h_{\mu\nu},
\end{eqnarray}

$a$ being some divergent constant.
Thus we set
$$P_{\mu\nu} = (\eta_{\mu\nu} + h_{\mu\nu}),
Q_{\mu\nu} = h_{\mu\nu}$$

at the zero order.
Then inclusion of the counterterms (\ref{cont1},\ref{cont2}) is just a multiplicative
renormalization of the source Y.

\vspace{0.5cm}

{\bf $\alpha$. One-loop order}

\vspace{0.5cm}

To obtain Slavnov identities at this order we perform a BRST-shift (\ref{brs})
of integration variables in (\ref{genernew})

\begin{eqnarray}\label{slavnov1}&&
{\displaystyle\int}dh_{\mu\nu}dC_{\sigma}d\bar{C}^{\tau}
\Big[ \left( T^{\mu\nu} + Y L^{\sigma} C^{\tau}\frac{\delta M^{(0)}_{\sigma\tau}}{\delta h_{\mu\nu}} + Y K^{\mu\nu} \right) \left(\frac{\delta}{\delta K^{\mu\nu}} + Y (\eta_{\mu\nu} + h_{\mu\nu}) \right)
\nonumber\\&&
- (\bar{\beta}_{\sigma} + Y L^{\tau} M^{(0)}_{\sigma\tau})\left( \frac{\delta}{\delta L^{\sigma}} + Y C^{\tau} M^{(0)}_{\sigma\tau} \right)
+ \frac{1}{\Delta}\beta_{\tau}\Box F^{\tau,\mu\nu}\frac{\delta}{\delta T^{\mu\nu}}
- 2 Y \Delta\frac{d}{d\Delta}
\nonumber\\&&
+ i Y \bar{C}^{\sigma}F_{\sigma}^{\mu\nu}D_{\mu\nu}^{\alpha}C_{\alpha}
\Big]
exp\{i (\tilde{\Sigma}
+ Y F_{\sigma}\bar{C}^{\sigma} + Y K^{\mu\nu}(\eta_{\mu\nu} + h_{\mu\nu})
+ Y L^{\sigma} C^{\tau} M^{(0)}_{\sigma\tau}
\nonumber\\&&
+ \bar{\beta}^{\sigma}C_{\sigma}
+ \bar{C}^{\tau}\beta_{\tau} + T^{\mu\nu}h_{\mu\nu})\} = 0.
\end{eqnarray}

Our aim is to find the $\Delta$-dependence of the gauge invariant
terms only. Terms containing $M^{(0)}_{\sigma\tau}$ in (\ref{slavnov1})
depending on anticommuting fields $C_{\sigma}$ and source $L^{\sigma}$
are unimportant in this respect and we replace them simply by " + ... "  in what follows
because these terms will be omitted in the end of the calculation anyway.

Using the ghost equation of motion

\begin{eqnarray}\label{ghosteqnew}&&
{\displaystyle\int}dh_{\mu\nu}dC_{\sigma}d\bar{C}^{\tau}
\left(F_{\tau}^{\mu\nu}D_{\mu\nu}^{\alpha}C_{\alpha}
- Y F_{\tau} + \beta_{\tau} \right)
\nonumber\\&&
exp\{i (\tilde{\Sigma} + Y F_{\sigma}\bar{C}^{\sigma}
+ Y K^{\mu\nu}(\eta_{\mu\nu} + h_{\mu\nu}) + ...
\nonumber\\&&
+ \bar{\beta}^{\sigma}C_{\sigma}
+ \bar{C}^{\tau}\beta_{\tau} + T^{\mu\nu}h_{\mu\nu})\} = 0,
\end{eqnarray}

introducing the generating functional of proper vertices $\tilde{\Gamma}$

\begin{eqnarray}&&
\tilde{\Gamma}[h_{\mu\nu},C_{\sigma},\bar{C}^{\tau},K^{\mu\nu},L^{\sigma},Y]
\nonumber\\&&
= W [T^{\mu\nu},\bar{\beta}^{\sigma},\beta_{\tau},K^{\mu\nu},L^{\sigma},Y]
-   \bar{\beta}^{\sigma}C_{\sigma} - \bar{C}^{\tau}\beta_{\tau} - T^{\mu\nu}h_{\mu\nu},
\end{eqnarray}

$W \equiv - i~lnZ,$

\begin{eqnarray}&&
h_{\mu\nu} = \frac{\delta W}{k\delta T^{\mu\nu}},
\nonumber\\&&
C_{\sigma} = \frac{\delta W}{\delta\bar{\beta}^{\sigma}},
\nonumber\\&&
\bar{C}^{\tau} = - \frac{\delta W}{\delta\beta_{\tau}}
\end{eqnarray}

and noting that

\begin{eqnarray}&&
\frac{d\tilde{\Gamma}}{d\Delta} = \frac{d W}{d\Delta},
\end{eqnarray}

we rewrite (\ref{slavnov1}) as the Slavnov identity for $\tilde{\Gamma}$

\begin{eqnarray}\label{slavnov1tg}&&
\frac{\delta\tilde{\Gamma}}{\delta h_{\mu\nu}}\left[ \frac{\delta\tilde{\Gamma}}{\delta K^{\mu\nu}} + Y (\eta_{\mu\nu} + h_{\mu\nu})\right]
+ \frac{\delta\tilde{\Gamma}}{\delta C_{\sigma}}\frac{\delta\tilde{\Gamma}}{\delta L^{\sigma}}
\nonumber\\&&
+ \frac{1}{\Delta}\Box F_{\tau}\frac{\delta\tilde{\Gamma}}{\delta\bar{C}_{\tau}}
+ 2 Y \Delta\frac{d\tilde{\Gamma}}{d\Delta}
+ Y \frac{\delta\tilde{\Gamma}}{\delta\bar{C}_{\tau}}\bar{C}_{\tau}
+ ... = 0.
\end{eqnarray}

To simplify eq.(\ref{slavnov1tg}) we introduce the reduced generating
functional

$${\Gamma} = \tilde{\Gamma} + \frac{1}{2\Delta}F_{\alpha}\Box F^{\alpha}
- Y K^{\mu\nu}(\eta_{\mu\nu} + h_{\mu\nu}) - Y F_{\sigma}\bar{C}^{\sigma}.$$

Then eq. (\ref{slavnov1tg}) reduces to

\begin{eqnarray}\label{slavnov1g}&&
\frac{\delta\Gamma}{\delta h_{\mu\nu}}\frac{\delta\Gamma}{\delta K^{\mu\nu}}
+ \frac{\delta\Gamma}{\delta C_{\sigma}}\frac{\delta\Gamma}{\delta L^{\sigma}}
\nonumber\\&&
+ 2 Y \Delta\frac{d\Gamma}{d\Delta}
+ Y K^{\mu\nu}\frac{\delta\Gamma}{\delta K^{\mu\nu}}
+ ... = 0.
\end{eqnarray}

The ghost equation of motion written in terms of $\Gamma$ is

\begin{eqnarray}\label{ghosteq1g}&&
F_{\tau}^{\mu\nu}\frac{\delta\Gamma}{\delta K^{\mu\nu}} - \frac{\delta\Gamma}{\delta\bar{C}^{\tau}} = 0.
\end{eqnarray}

Now let us separate Y-independent part of $\Gamma$ from the part linear in Y

\begin{eqnarray}\label{sep}&&
\Gamma = \Gamma_1 + Y\Gamma_2.
\end{eqnarray}

Then eq.(\ref{slavnov1g}) gives an ordinary Slavnov identity for $\Gamma_1$

\begin{eqnarray}\label{slavord1}&&
\frac{\delta\Gamma_1}{\delta h_{\mu\nu}}\frac{\delta\Gamma_1}{\delta K^{\mu\nu}}
+ \frac{\delta\Gamma_1}{\delta C_{\sigma}}\frac{\delta\Gamma_1}{\delta L^{\sigma}}
= 0
\end{eqnarray}

and an equation involving $\Gamma_2$

\begin{eqnarray}&&
- \frac{\delta\Gamma_1}{\delta h_{\mu\nu}}\frac{\delta\Gamma_2}{\delta K^{\mu\nu}}
+ \frac{\delta\Gamma_2}{\delta h_{\mu\nu}}\frac{\delta\Gamma_1}{\delta K^{\mu\nu}}
+ \frac{\delta\Gamma_1}{\delta C_{\sigma}}\frac{\delta\Gamma_2}{\delta L^{\sigma}}
\nonumber\\&&
- \frac{\delta\Gamma_2}{\delta C_{\sigma}}\frac{\delta\Gamma_1}{\delta L^{\sigma}}
+ 2\Delta\frac{d\Gamma_1}{d\Delta}
+ K^{\mu\nu}\frac{\delta\Gamma_1}{\delta K^{\mu\nu}} + ... = 0.
\end{eqnarray}

Finally, we omit all but the terms depending on $h_{\mu\nu}$ only and obtain in the first order

\begin{eqnarray}&&
- \frac{\delta S_{0}}{\delta h_{\mu\nu}}\frac{\delta\Gamma_2^{div(1)}}{\delta K^{\mu\nu}}
+ 2\Delta\frac{d\Omega^{div(1)}}{d\Delta} + ... = 0,
\end{eqnarray}

where $\Omega$ denotes the gauge invariant part of $\Gamma_1$ and the superscript
$div(1)$ denotes the one-loop divergent part of the corresponding quantities.

As we know $\Gamma_2^{div(1)} = K^{\mu\nu}P^{(1)}_{\mu\nu}$,
$P^{(1)}$ being some divergent polynom in $h_{\mu\nu}.$

Thus, dropping out the terms proportional to $K^{\mu\nu}$ again and the symbol " + ... "
we obtain the following equation for the gauge invariant terms
$\Omega^{div(1)}$ of the effective action\footnote{We will see in secs.3,4 that the non-gauge-invariant terms in $\Gamma_1^{div}$ depending on $h_{\mu\nu}$ only are absent.}

\begin{eqnarray}\label{res1}&&
2\Delta\frac{d\Omega^{div(1)}}{d\Delta} =
\frac{\delta S_{0}}{\delta h_{\mu\nu}}P^{(1)}_{\mu\nu}.
\end{eqnarray}

The left hand side of this equation is gauge invariant thus so is the right hand side.
Therefore $P_{\mu\nu}^{(1)}$ has the form mentioned above.

To make the Green functions finite at the one-loop level we must redefine
the initial effective action $\Sigma$

\begin{eqnarray}\label{redef}&&
\Sigma \to \Sigma^{(1)} = \Sigma - \Gamma_1^{div(1)}
\end{eqnarray}

and the source Y\footnote{We should also include counterterms
of the type YLC, but they are irrelevant to the issue and replaced by " + ... " as we have mentioned above.}

\begin{eqnarray}\label{redefy}&&
Y \to Y (1 - a^{(1)} ).
\end{eqnarray}

As explained in \cite{stelle} subtraction of $\Gamma_1^{div(1)}$ boils down to
a redefinition of all the fields in such a way that $\Sigma^{(1)}$ is invariant
under renormalized set of BRST-transformations for which we do not introduce new notation.

\vspace{0.5cm}

{\bf $\beta$. Two-loop order}

\vspace{0.5cm}

We perform a renormalized BRST-transformation of integration variables in the
generating functional of Green functions finite at one-loop level

\begin{eqnarray}\label{generoneloop}&&
Z^{[1]}[T^{\mu\nu},\bar{\beta}^{\sigma},\beta_{\tau},K^{\mu\nu},L^{\sigma},Y] =
\nonumber\\&&
{\displaystyle\int}dh_{\mu\nu}dC_{\sigma}d\bar{C}^{\tau}
exp\{i (\tilde{\Sigma}^{(1)}
+ Y (1 - a^{(1)}) F_{\sigma}\bar{C}^{\sigma} + (1 - a^{(1)}) Y K^{\mu\nu} (\eta_{\mu\nu} + h_{\mu\nu})
\nonumber\\&&
+ ... + \bar{\beta}^{\sigma}C_{\sigma}
+ \bar{C}^{\tau}\beta_{\tau} + T^{\mu\nu}h_{\mu\nu})\}
\end{eqnarray}

and obtain the following Slavnov identity

\begin{eqnarray}\label{slavnov2}&&
{\displaystyle\int}dh_{\mu\nu}dC_{\sigma}d\bar{C}^{\tau}
\Big[ \left( T^{\mu\nu} + Y (1 - a^{(1)}) K^{\mu\nu} \right) \left(\frac{\delta}{\delta K^{\mu\nu}} + Y (1 - a^{(1)}) (\eta_{\mu\nu} + h_{\mu\nu}) \right)
\nonumber\\&&
- \bar{\beta}^{\sigma}\frac{\delta}{\delta L^{\sigma}}
+ \frac{1}{\Delta}\beta_{\tau}\Box F^{\tau,\mu\nu}\frac{\delta}{\delta T^{\mu\nu}}
- 2 Y (1 - a^{(1)}) \Delta\left(\frac{d}{d\Delta} + \frac{d \Gamma_{1}^{div(1)}}{d\Delta}  \right)
\nonumber\\&&
+ i Y (1 - a^{(1)}) \bar{C}^{\sigma}F_{\sigma}^{\mu\nu}D_{\mu\nu}^{\alpha}C_{\alpha} + ...
\Big]
exp\{i (\tilde{\Sigma}^{(1)}
+ Y (1 - a^{(1)}) F_{\sigma}\bar{C}^{\sigma}
\nonumber\\&&
+ Y (1 - a^{(1)}) K^{\mu\nu}(\eta_{\mu\nu} + h_{\mu\nu}) + ...
+ \bar{\beta}^{\sigma}C_{\sigma}
+ \bar{C}^{\tau}\beta_{\tau} + T^{\mu\nu}h_{\mu\nu})\} = 0.
\end{eqnarray}

To evaluate the term\footnote{Again evaluation of the gauge invariant part
of $\Delta\frac{d\Gamma_1^{div(1)}}{d\Delta}$ is needed only.}

\begin{eqnarray}&&
{\displaystyle\int}dh_{\mu\nu}dC_{\sigma}d\bar{C}^{\tau}
Y\Delta\frac{d\Gamma_1^{div(1)}}{d\Delta}
exp\{i (\tilde{\Sigma}^{(1)}
+ Y ( 1 - a^{(1)}) F_{\sigma}\bar{C}^{\sigma}
\nonumber\\&&
+ ( 1 - a^{(1)}) Y K^{\mu\nu}(\eta_{\mu\nu} + h_{\mu\nu})
+ ... + \bar{\beta}^{\sigma}C_{\sigma}
+ \bar{C}^{\tau}\beta_{\tau} + T^{\mu\nu}h_{\mu\nu})\}
\end{eqnarray}

we use eq.(\ref{res1}) and equation of motion of h-field
which is obtained from (\ref{generoneloop})\footnote{We use the property $Y^2 = 0.$}

\begin{eqnarray}&&
Y {\displaystyle\int}dh_{\mu\nu}dC_{\sigma}d\bar{C}^{\tau}
\left(\frac{\delta\tilde{\Sigma}}{\delta h_{\mu\nu}}
- \frac{\delta\Gamma_1^{div(1)}}{\delta h_{\mu\nu}}
+ T^{\mu\nu} \right)
\nonumber\\&&
exp\{i (\tilde{\Sigma}^{(1)}
+ Y (1 - a^{(1)}) F_{\sigma}\bar{C}^{\sigma}
\nonumber\\&&
+ Y (1 - a^{(1)}) K^{\mu\nu}(\eta_{\mu\nu} + h_{\mu\nu}) + ...
+ \bar{\beta}^{\sigma}C_{\sigma}
+ \bar{C}^{\tau}\beta_{\tau} + T^{\mu\nu}h_{\mu\nu})\} = 0.
\end{eqnarray}

In the two-loop approximation we may write

\begin{eqnarray}&&
a^{(1)}{\displaystyle\int}dh_{\mu\nu}dC_{\sigma}d\bar{C}^{\tau}
\frac{\delta\Gamma_1^{div(1)}}{\delta h_{\mu\nu}}exp\{i...\}
= a^{(1)}\frac{\delta\Gamma_1^{div(1)}}{\delta h_{\mu\nu}} Z^{[1]}.
\end{eqnarray}

Finally, using the ghost equation of motion

\begin{eqnarray}\label{ghosteq2g}&&
F_{\tau}^{\mu\nu}\frac{\delta\Gamma^{[1]}}{\delta K^{\mu\nu}} - \frac{\delta\Gamma^{[1]}}{\delta\bar{C}^{\tau}} = 0,
\end{eqnarray}

written in terms of the one-loop finite reduced generating functional of proper
vertices

$$\Gamma^{[1]} = \tilde{\Gamma}^{[1]} + \frac{1}{2\Delta}F_{\alpha}\Box F^{\alpha}
- Y K^{\mu\nu}(\eta_{\mu\nu} + h_{\mu\nu}) - Y F_{\sigma}\bar{C}^{\sigma}.$$

we rewrite the rest of the eq.(\ref{slavnov2})
as in section $\alpha$ and obtain the following Slavnov identity for one-loop finite proper vertices,
valid up to two-loop order

\begin{eqnarray}\label{slavnov2g}&&
\frac{\delta\Gamma^{[1]}}{\delta h_{\mu\nu}}\frac{\delta\Gamma^{[1]}}{\delta K^{\mu\nu}}
+ \frac{\delta\Gamma^{[1]}}{\delta C_{\sigma}}\frac{\delta\Gamma^{[1]}}{\delta L^{\sigma}}
+ 2 Y \Delta\frac{d\Gamma^{[1]}}{d\Delta}
+ Y a^{(1)} \frac{\delta\Gamma_1^{div(1)}}{\delta h_{\mu\nu}}(\eta_{\mu\nu} + h_{\mu\nu})
+ ... = 0,
\end{eqnarray}

where terms explicitly dependent on $K^{\mu\nu}$ and $\bar{C}_{\sigma}$ are included in " + ... " for simplicity.

Again the separation

\begin{eqnarray}\label{sep2}&&
\Gamma^{[1]} = \Gamma_1^{[1]} + Y\Gamma_2^{[1]}
\end{eqnarray}

gives an ordinary Slavnov identity

\begin{eqnarray}\label{slavord2}&&
\frac{\delta\Gamma_1^{[1]}}{\delta h_{\mu\nu}}\frac{\delta\Gamma_1^{[1]}}{\delta K^{\mu\nu}}
+ \frac{\delta\Gamma_1^{[1]}}{\delta C_{\sigma}}\frac{\delta\Gamma_1^{[1]}}{\delta L^{\sigma}}
= 0
\end{eqnarray}

and an identity involving $\Gamma_2^{{[1]}}$

\begin{eqnarray}&&
- \frac{\delta\Gamma_1^{[1]}}{\delta h_{\mu\nu}}\frac{\delta\Gamma_2^{[1]}}{\delta K^{\mu\nu}}
+ \frac{\delta\Gamma_2^{[1]}}{\delta h_{\mu\nu}}\frac{\delta\Gamma_1^{[1]}}{\delta K^{\mu\nu}}
+ \frac{\delta\Gamma_1^{[1]}}{\delta C_{\sigma}}\frac{\delta\Gamma_2^{[1]}}{\delta L^{\sigma}}
- \frac{\delta\Gamma_2^{[1]}}{\delta C_{\sigma}}\frac{\delta\Gamma_1^{[1]}}{\delta L^{\sigma}}
\nonumber\\&&
+ 2 \Delta\frac{d\Gamma_1^{[1]}}{d\Delta}
+ a^{(1)}\frac{\delta \Gamma_1^{div(1)}}{\delta h_{\mu\nu}}(\eta_{\mu\nu} + h_{\mu\nu})
+ ... = 0.
\end{eqnarray}

Thus for two-loop gauge invariant divergent part $\Omega^{[1]div(2)}$ of
the one-loop finite generating functional of proper vertices we have

\begin{eqnarray}\label{main}&&
- \frac{\delta S_{0}}{\delta h_{\mu\nu}} P_{\mu\nu}^{(2)}
+ 2\Delta\frac{d\Omega^{[1]div(2)}}{d\Delta}
\nonumber\\&&
+ a^{(1)}\frac{\delta\Omega^{div(1)}}{\delta h_{\mu\nu}}(\eta_{\mu\nu} + h_{\mu\nu}) = 0.
\end{eqnarray}

Again it follows from eq.(\ref{main}) that $P_{\mu\nu}^{(2)} = a^{(2)} (\eta_{\mu\nu} + h_{\mu\nu}).$

Thus we obtain the following identity for one- and two-loop
divergent gauge invariant parts of the effective action

\begin{eqnarray}\label{res2}&&
2\Delta\frac{d\Omega^{[1]div(2)}}{d\Delta} =
- a^{(1)}\frac{\delta\Omega^{div(1)}}{\delta h_{\mu\nu}}(\eta_{\mu\nu} + h_{\mu\nu})
+ a^{(2)}\frac{\delta S_{0}}{\delta h_{\mu\nu}} (\eta_{\mu\nu} + h_{\mu\nu}).
\end{eqnarray}

Had we used any other parametrization of the gravitational field, $P^{(1)}$ and
$P^{(2)}$ would have such a form that provides the gauge invariance of
the product $\frac{\delta S_{0}}{\delta h_{\mu\nu}} P_{\mu\nu}$,
where $h_{\mu\nu}$ denotes the set of standard variables.
Therefore result (\ref{res2}) holds in general if $h_{\mu\nu}$
denotes a quantum part of the covariant components of the metric field.

Thus we see that in the presence of the new source Y the renormalization
procedure differs from the usual one substantially.
Although the renormalized Green functions satisfy the same Slavnov
identities as the bare ones, the renormalization equation (of the type (\ref{res2}))
for their divergent parts in (n+1)-th loop order can not be obtained by
simple omitting of the finite parts of the Slavnov identities for the
Green functions renormalized up to n-th loop order.
The correct procedure presented above leads to the Slavnov identities
which just impose some nontrivial constraints on the form of gauge-dependent
divergent structures of the Green functions.

\vspace{1cm}

{\large{\bf 3. Calculation of the one-loop divergent part of $\Omega$}}

\vspace{1cm}

In previous section we have obtained the relation (\ref{res2})
which identifies ({\it modulo} terms proportional to the equations
of motion of h field) the $\Delta$-derivatives of the two-loop
gauge-invariant divergent part of the effective action with the
variational derivatives of the corresponding one-loop part up to
some coefficient being defined by divergent parts of diagrams with one
insertion of the Y vertex. To prove this coefficient is not zero we present
explicit calculation of the values $\Gamma_1^{div(1)}$ and $\Gamma_1^{[1]div(2)}$
in arbitrary gauge of the type (\ref{gauge}) and arbitrary parametrization
with the only restriction of linearity of group generators. We prefer
this way to direct computation of diagrams with Y insertion because
it allows to verify the relation (\ref{res2}).

\vspace{0,5cm}

{\bf a. Arbitrary parametrizations}

\vspace{0,5cm}

In general the metric is an arbitrary function of dynamical variables.
The only restriction is that this function must be nondegenerate.
For example, if dynamical variables are chosen as $g^{*}_{\mu\nu} = g_{\mu\nu} ( - g )^{r},
g = det g_{\mu\nu}$ then we should avoid the case of $r = - \frac{1}{4}$;
otherwise $det g^{*}_{\mu\nu} = 1$ and one more independent variable must be
introduced in addition to the set of $g^{*}_{\mu\nu}$

To calculate one-loop divergences the background field method is used \cite{dewitt,hooft,hooftvelt}.
Accordingly, we should first find expansions of all the quantities entering (\ref{eqofmot})
in powers of dynamical quantum variables $h_{\mu\nu}$ around the background
field $g_{\mu\nu}$ up to the second order.

Now we note that the form of the graviton propagator is, of course, parametrization-dependent
and this dependence complicates all calculations considerably. However,
it is fictitious in the sense that always can be removed by {\it linear} redefinition
of quantum variables. Such a change doesn't mix different orders in the loop
expansion and therefore doesn't alter the values of the one-loop divergent
part in particular\footnote{The corresponding Jacobian is $exp ( \delta (0)...) = 1$ in the dimensional regularization.}.
We will show below that our calculations are highly simplified if the linear
part of the metric expansion is chosen to have the simplest form $h_{\mu\nu}$

\begin{equation}\label{standard}
\underline{g}_{\mu\nu} = g_{\mu\nu} + h_{\mu\nu} + a h h_{\mu\nu} +
b h_{\mu\alpha} h^{\alpha}_{\nu} + c h^2 g_{\mu\nu}
+ d h_{\alpha\beta} h^{\alpha\beta} g_{\mu\nu}+O(h^3),
\end{equation}

where $a, b, c, d$ are arbitrary constants; $\underline{g}_{\mu\nu}$ denotes
the full metric field; all raising of indices is done by means of the
inverse background metric $g^{\mu\nu}$: $g_{\mu\alpha} g^{\alpha\nu} = \delta^{\nu}_{\mu};$
$h \equiv h_{\mu\nu} g^{\mu\nu}.$ Any parametrization $g^{*}_{\mu\nu}$ somehow
constructed from the metric $g_{\mu\nu}$ and its determinant has a background
expansion reducible to (\ref{standard}).

To show an advantage of such a choice of the background metric expansion we note
that it is only the linear part of this expansion which in fact contributes to the
curvature $R^{\mu}_{{}\nu\alpha\beta}$ expansion. Really, this tensor
has the following structure

\begin{equation}
\underline{R} = \partial\underline{\Gamma} - \partial\underline{\Gamma} +
{\underline{\Gamma}} \underline{\Gamma} - {\underline{\Gamma}} \underline{\Gamma}.
\end{equation}

Suppose at the moment that we have chosen our coordinate system in such a way
that $\Gamma$ = 0 at any fixed point of space-time.\footnote{Recall that $\underline{\Gamma}$ and $\Gamma$ are constructed from $\underline{g}_{\mu\nu}$ and $g_{\mu\nu}$ respectively.}
Then

\begin{equation}
\underline{R} = \partial\Gamma - \partial\Gamma + \Gamma_1\Gamma_1 - \Gamma_1\Gamma_1 + \partial (\Gamma_1 + \Gamma_2 ) - \partial (\Gamma_1 + \Gamma_2) + O(h^3),
\end{equation}

where subscripts 1 and 2 denote parts of $\underline{\Gamma}$ of the
corresponding powers in h.
We may rewrite this expression in explicitly covariant form as

\begin{equation}
\underline{R} = R + \nabla ( \Gamma_1 + \Gamma_2 ) - \nabla ( \Gamma_1 + \Gamma_2 ) + \Gamma_1\Gamma_1 - \Gamma_1\Gamma_1 + O(h^3), \footnote{Note that $\Gamma_{1,2}$ are tensors.}
\end{equation}

valid therefore in every coordinate system. In terms of the Lagrangian linear
in curvature scalar all the full derivatives of the second order may be dropped
out. In quadratic terms these derivatives are multiplied by the zeroth order
quantities $R_{\mu\nu}$, $R$ etc., when the second variation of the action is
being calculated. Integrating by parts one can easily verify by means
of power counting that these terms do not contribute to the one-loop
divergent part of the effective action.

It follows from the above discussion that the dependence on the
parametrization appears only in terms without h-derivatives in the second
variation of the action if the reduced expansion (\ref{standard}) is used.

\vspace{0,5cm}

{\bf b. One-loop invariance on-shell.}

\vspace{0,5cm}

To calculate one-loop divergent part of the effective action in arbitrary
gauge and parametrization we shall use the fact that the dependence
on parameters $\Delta$, $a, b, c, d$ appears in the terms proportional
to the equations of motion only. As far as the $\Delta$-dependence is concerned
the corresponding result follows directly from eq.(\ref{res1}).

To prove the on-shell independence from $a, b, c, d$ we note first of all
that these parameters appear in the second variation of the action only
in terms having the form

$$\int \frac{\delta S}{\delta g_{\mu\nu}} [g^*_{\mu\nu}]_2,$$

where $[g^{*}_{\mu\nu}]_2$ denotes the second order part of the reduced
metric expansion (\ref{standard}).

Next, calculating generators of the gauge transformations of dynamical
variables belonging to the class of parametrizations described above
and passing to the set of standard variables again one easily sees that
these generators just coincide with the ordinary ones of the metric field
transformations, i.e. they are $a, b, c, d$ - independent and therefore
so is the ghost contribution.

Thus the on-shell invariance is proved.

\vspace{0,5cm}

{\bf c. Background field method.}

\vspace{0,5cm}

According to the background field method we separate the quantum field part $h^{*}_{\mu\nu}$
>from the external field $g^{*}_{\mu\nu}$

$$\underline{g}^{*}_{\mu\nu} = g^{*}_{\mu\nu} + h^{*}_{\mu\nu} .$$

Then we expand the metric field $g_{\mu\nu}$ in powers of $h^{*}_{\mu\nu}$
and perform a linear transformation on $h^{*}_{\mu\nu}$ bringing this
expansion to the form of (\ref{standard}).

Imposing the background Lorentz gauge on the quantum field $h_{\mu\nu}$

\begin{eqnarray}&&
F_{\mu}(g)\equiv F^{\alpha\beta}_{\mu}(g)h_{\alpha\beta}\equiv \nabla^{\nu}h_{\mu\nu},
\end{eqnarray}

we have for the generating functional of Green functions

\begin{eqnarray}\label{generback}&&
Z[T^{\mu\nu}] =
{\displaystyle\int}dh_{\mu\nu}dC_{\sigma}d\bar{C}^{\tau} \left\{ det~g_{\mu\nu}\nabla^2 \right\}^{\frac{1}{2}}
\nonumber\\&&
exp\{i (S_{0}(g,h) - \frac{1}{2\Delta}F^{\alpha}(g)\nabla^2 F_{\alpha}(g)\sqrt{-g}
+ \bar{C}^{\tau}F_{\tau}^{\mu\nu}(g)D^{\alpha}_{\mu\nu}C_{\alpha}
+ T^{\mu\nu}h_{\mu\nu})\}.
\end{eqnarray}

We suppose that the background field $g_{\mu\nu} - \eta_{\mu\nu}$ and the source $T^{\mu\nu}$ are absent out of
some finite region of space-time. Integration is carried out in all fields
$h_{\mu\nu}$ tending to zero at infinity.
We do not introduce background ghost fields nor their sources because
renormalization of these fields plays no role in this section nor in sec.4.

In the one-loop approximation we expand the gauge fixed action
$$ S_{gf}\equiv S_{0}(g,h) - \frac{1}{2\Delta}F^{\alpha}(g)\nabla^2 F_{\alpha}(g)\sqrt{-g}$$
around the extremal $\tilde{h}$ satisfying the classical equations of motion

\begin{eqnarray}\label{class}
\frac{\delta S_{gf}(g,h)}{\delta h_{\mu\nu}} + T^{\mu\nu} = 0
\end{eqnarray}

up to the second order and obtain

\begin{eqnarray}\label{generbackoneloop}&&
Z[T^{\mu\nu}] =
exp\{i (S_{gf}(g,\tilde{h})+ T^{\mu\nu}\tilde{h}^{\mu\nu})\}
\left\{ det~g_{\mu\nu}\nabla^2\right\}^{\frac{1}{2}}
\nonumber\\&&
{\displaystyle\int}dh_{\mu\nu}dC_{\sigma}d\bar{C}^{\tau}
det F^{\mu\nu}_{\beta}(g)D^{\alpha}_{\mu\nu}(g,h)
exp\left\{\frac{i}{2} \frac{\delta^2 S_{gf}(g,\tilde{h})}{\delta h_{\mu\nu}\delta h_{\alpha\beta}} ( h_{\mu\nu} - \tilde{h}_{\mu\nu} )( h_{\alpha\beta} - \tilde{h}_{\alpha\beta})\right\}.
\end{eqnarray}

As far as we have supposed the background field $g - \eta$ and the source $T$ to disappear
out of some finite region of space-time one can choose a solution
$\tilde{h}$ of eq.(\ref{class}) to be zero at infinity. Thus the shift of
integration variables $h \to h + \tilde{h}$ doesn't change boundary
conditions for $h$ and we have for the generating functional of connected
Green functions

\begin{eqnarray}&&
W \equiv - i ~ln Z =
S_{gf}(g,\tilde{h}) + T^{\mu\nu}\tilde{h}_{\mu\nu} +
\frac{i}{2} ~Tr ~ln \frac{\delta^2 S_{gf}(g,\tilde{h})}{\delta h_{\mu\nu}\delta h_{\alpha\beta}} -
\nonumber\\&&
i ~Tr ~ln ~F^{\mu\nu}_{\tau}(g)D^{\alpha}_{\mu\nu}(g,\tilde{h}) - \frac{i}{2}~Tr~ln~g_{\mu\nu}\nabla^2.
\end{eqnarray}

To perform a Legendre transformation we calculate

\begin{eqnarray}&&
h_{\mu\nu} \equiv \frac{\delta W}{\delta T^{\mu\nu}} =
\nonumber\\&&
\tilde{h}_{\mu\nu}
+ \frac{\delta}{\delta T^{\mu\nu}}\left\{
\frac{i}{2} ~Tr ~ln \frac{\delta^2 S_{gf}(g,\tilde{h})}{\delta h_{\mu\nu}\delta h_{\alpha\beta}} -
i ~Tr ~ln ~F^{\mu\nu}_{\tau}(g)D^{\alpha}_{\mu\nu}(g,\tilde{h})
\right\}
\end{eqnarray}

and

\begin{eqnarray}&&
\Gamma(g,h) = W(g,h) - h_{\mu\nu}T^{\mu\nu} =
S_{gf}(g,\tilde{h}) +
\frac{i}{2} ~Tr ~ln \frac{\delta^2 S_{gf}(g,\tilde{h})}{\delta h_{\mu\nu}\delta h_{\alpha\beta}} -
\nonumber\\&&
i ~Tr ~ln ~F^{\mu\nu}_{\tau}(g)D^{\alpha}_{\mu\nu}(g,\tilde{h}) -
\frac{i}{2}~Tr~ln~g_{\mu\nu}\nabla^2 -
\nonumber\\&&
T^{\mu\nu}\frac{\delta}{\delta T^{\mu\nu}}
\left\{\frac{i}{2} ~Tr ~ln \frac{\delta^2 S_{gf}(g,\tilde{h})}{\delta h_{\mu\nu}\delta h_{\alpha\beta}} -
i ~Tr ~ln ~F^{\mu\nu}_{\tau}(g)D^{\alpha}_{\mu\nu}(g,\tilde{h})
\right\} =
S_{gf}(g,h) +
\nonumber\\&&
\frac{i}{2} ~Tr ~ln \frac{\delta^2 S_{gf}(g,h)}{\delta h_{\mu\nu}\delta h_{\alpha\beta}} -
i ~Tr ~ln ~F^{\mu\nu}_{\tau}(g)D^{\alpha}_{\mu\nu}(g,h)
- \frac{i}{2}~Tr~ln~g_{\mu\nu}\nabla^2;
\end{eqnarray}

the eq.(\ref{class}) was used in the last passage.

Obtaining the relation (\ref{res2}) in sec.2 we used the flat background $\eta_{\mu\nu}$.
Had we started with an arbitrary background metric $g_{\mu\nu}$ instead of $\eta_{\mu\nu}$
we would have modulo terms proportional to the equations of motion

\begin{eqnarray}\label{res2b}&&
2\Delta\frac{d\Omega^{[1]div(2)}}{d\Delta} =
- a^{(1)}\frac{\delta\Omega^{div(1)}}{\delta g_{\mu\nu}} g_{\mu\nu}.
\end{eqnarray}

We wrote $g_{\mu\nu}$ instead of $\underline{g}_{\mu\nu}$ in (\ref{res2b})
because it is sufficient to verify this relation in the case $h_{\mu\nu} = 0.$

\vspace{0,5cm}

{\bf d. Calculation of $\Omega^{div(1)}$.}

\vspace{0,5cm}

Let us first reveal some "essential" properties of charges.

As $R^2$-gravity is renormalizable we can write the $\Omega^{div(1)}$ in the form

\begin{eqnarray}\label{r2gen}
&&\hspace{-1,5cm}
\Omega^{div(1)} =
\frac{1}{32\pi^2\varepsilon} \int d^4 x\ \sqrt{-g}
\left(c_1 R^{\mu\nu} R_{\mu\nu} + c_2 R^2 + c_3 R + c_4 \Lambda + c_5 \right),
\end{eqnarray}

where $c_i, i = 1,...,5$ are some
gauge and parametrization dependent coefficients.

As we know from sec.b. $\Omega^{div(1)}$ is gauge and parametrization independent on shell.
It is obvious that the only scalar which can be constructed from equations (\ref{eqmot})
to transform $\Omega^{div(1)}$ (\ref{r2gen}) is

\begin{eqnarray}\label{eqmot1}&&
\frac{1}{k^2}(R - 4 \Lambda) = - 2 (3 \alpha_1 + \beta)\nabla^2 R.
\end{eqnarray}

It follows from these simple facts that $c_i, i = 1,2,5$ and the combination
$4 c_3 + c_4$ do not depend on $\Delta, a, b, c, d.$

Thus we may simplify the calculation of $\Omega^{div(1)}$ in arbitrary gauge
and parametrization if divide it into two parts:

1. Calculation of $\Omega^{div(1)}$ in the case of the simplest gauge and
parametrization. We choose $g^{*}_{\mu\nu} = g_{\mu\nu}$ and the minimal gauge.

2. Calculation of the coefficient $c_4$ alone in arbitrary gauge and parametrization.
In this part we may obviously consider the space-time as flat.

The correct result of the first part of our program was obtained in \cite{avrbar}

\begin{eqnarray}\label{answerR2}
&&\hspace{-1,5cm}
\Omega^{div(1)} =
\frac{1}{32\pi^2\varepsilon} \int d^4 x\ \sqrt{-g}
\left(c_1 R^{\mu\nu} R_{\mu\nu} + c_2 R^2 + c_3 R + c_4 \Lambda + c_5 \right),
\end{eqnarray}

where

$$c_1 = \frac{133}{10},
c_2 = \frac{10\alpha_1^2}{\beta^2} + \frac{10\alpha_1}{6\beta} - \frac{291}{60},
c_3 = - \frac{1}{ (3  \alpha_1 + \beta) k^2}\left[  \frac{30 \alpha_1^2}{\beta^2} + \frac{53\alpha_1}{2\beta} + \frac{21}{4} \right],$$

$$c_4 = \frac{1}{(3  \alpha_1 + \beta) k^2} \left[ \frac{28 \alpha_1}{\beta} + 9 \right],
c_5 = \frac{3}{(3  \alpha_1 + \beta)^2 k^4}  \left[ \frac{15\alpha_1^2}{2\beta^2} + \frac{5\alpha_1}{\beta} + \frac{7}{8} \right].$$

Calculation of $c_4$ in the flat space-time is presented in Appendix A.
Combination of the two results gives

\begin{eqnarray}\label{answerr2gen}
&&\hspace{-1,5cm}
\Omega^{div(1)} =
\frac{1}{32\pi^2\varepsilon} \int d^4 x\ \sqrt{-g}
\left(c_1 R^{\mu\nu} R_{\mu\nu} + c_2 R^2 + c_3 R + c_4 \Lambda + c_5 \right),
\end{eqnarray}

where

$$c_1 = \frac{133}{10},\hspace{.5cm} c_2 = \frac{10}{9} \alpha^2 - \frac{5}{3}\alpha - \frac{773}{180}, \hspace{.5cm}
c_5 = \frac{1}{\beta^2 k^4}\left(\frac{5}{2} + \frac{1}{8\alpha^2}\right),$$

\begin{eqnarray}
c_4 &=&
- \frac{1}{\beta k^2}\left[ u \left( 2\Delta + \frac{3}{\alpha} \right) + v \left( 14 \Delta + \frac{1}{\alpha} + 20 \right) \right],
\nonumber
\end{eqnarray}

$$4 c_3 + c_4 = \frac{1}{\beta k^2}\left[ \frac{1}{3 \alpha} + 10 - \frac{40 \alpha}{3} \right],
\alpha = \frac{3 \alpha_1}{\beta} + 1, u = a + 4 c + \frac{1}{4}, v = b + 4 d - \frac{1}{2}.$$

\pagebreak

\vspace{1cm}

{\large{\bf 4. Calculation of the two-loop divergent part of $\Omega$}}

\vspace{1cm}

As follows from eq.(\ref{res2}) the nontrivial dependence on the gauge
parameter $\Delta$ (i.e. which is not zero modulo equations of motion)
is contained in terms proportional to $\frac{1}{\varepsilon^2}$.
To calculate the latter we use the renormgroup method.
It is very convenient to apply the generalized version of the
renormgroup equations given in \cite{blox,kaz}. For the sake of completeness
we give an account of this method following \cite{kaz}.

\vspace{0,5cm}

{\bf a. Generalized renormgroup method}

\vspace{0,5cm}

The idea of this approach is to obtain renormgroup equations without
explicit distinguishing of different charges, i.e. in terms
of the whole Lagrangian.

Let us consider the bare Lagrangian $L^{b}$ as a functional of the initial
Lagrangian L

\begin{eqnarray} \label{bare}
L^b = (\mu^2){}^\varepsilon \left\{L + \sum_{n = 1}^{\infty} \frac{A_n(L)}{\varepsilon^n}\right\},
\end{eqnarray}

where symbol $A_{n}(L)$ means that the corresponding counterpart
is calculated for the Lagrangian L. Independence of the $L^{b}$
>from the mass scale implies

\begin{eqnarray}\label{beta}
\beta(L) = \left(L\frac{\delta}{\delta L} - 1\right) A_1(L),
\end{eqnarray}
\begin{eqnarray}\label{recurr}
\left(L\frac{\delta}{\delta L} - 1\right)A_n(L) = \beta(L)\frac{\delta}{\delta L}A_{n - 1}(L),
\end{eqnarray}

where the so called generalized $\beta$-function $\beta(L)$ is defined by

$$\left.\mu^2 \frac{d L}{d \mu^2}\right|_{L^b} = - \varepsilon L + \beta (L),$$

We don't have to muse upon the concrete sense which the operation
$\frac{\delta}{\delta L}$ possesses. Using the loop expansion
of $A_{n}$

$$A_n(L) = \sum_{k = n}^{\infty} A_{nk}(L),$$

and noting the homogeneity of functionals $A_{nk}(L)$

$$A_{nk}(\lambda L) = \lambda^{1 - k}A_{nk}(L),$$
($\lambda$ being a constant)

we can express the operations $L \frac{\delta}{\delta L}$ and
$\beta (L)\frac{\delta}{\delta L}$ in terms of the ordinary
differentiation

\begin{eqnarray}
L\frac{\delta}{\delta L} A_{nk}(L) = \left.\frac{\partial}{\partial\lambda}A_{nk}(\lambda L)\right|_{\lambda = 1} = ( 1 - k )A_{nk}(L)
\end{eqnarray}

and

\begin{eqnarray}
\beta (L)\frac{\delta}{\delta L} A_{nk}(L) = \left.\frac{d}{d x} A_{nk}(L + x \beta (L))\right|_{ x = 0 },
\end{eqnarray}

whatever meaning has to be assigned to $\frac{\delta}{\delta L}$.

Thus we obtain

\begin{eqnarray}
\beta(L) = \left.\left(\frac{\partial}{\partial \lambda} - 1\right)A_1(\lambda L)\right|_{\lambda = 1} = \sum_{k = 1}^{\infty} - k A_{1k}(L),
\end{eqnarray}

\begin{eqnarray}\label{sv}
\left.\left(\frac{\partial}{\partial\lambda} - 1\right)\sum_{k = n}^{\infty} A_{nk}(\lambda L)\right|_{\lambda = 1} = \left.\frac{d}{d x} \sum_{k = n - 1}^{\infty} A_{n - 1,k}\left(L - x \sum_{l = 1}^{\infty}l A_{1l}(L)\right)\right|_{x = 0}.
\end{eqnarray}

To relate $A_{nn}$ and $A_{n - 1,n - 1}$ we substitute $L \to \xi^{-1} L$ in (\ref{sv}),
differentiate with respect to $\xi$ $n - 1$ times and set $\xi$ = 0.

The result is

\begin{eqnarray}
n A_{nn}(L) = \left.\frac{d}{d x} A_{n - 1,n - 1}(L + x A_{11}(L))\right|_{x = 0}.
\end{eqnarray}

\vspace{1cm}

{\bf b. Calculation of $\Omega_{1/\varepsilon^2}^{[1]div(2)}$}

\vspace{1cm}

To apply the relation

\begin{eqnarray}\label{22-11}
A_{22}(L) = \frac{1}{2}\left.\frac{d}{d x}A_{11}(L + x A_{11}(L))\right|_{x = 0}.
\end{eqnarray}

to the case of

\begin{eqnarray}&&
L_{eff} = \sqrt{-g}(\alpha_1 R^2 + \beta R_{\mu\nu} R^{\mu\nu}
- \frac{1}{k^2} (R - 2 \Lambda))
\nonumber\\&&
- \frac{1}{2\Delta}F_{\alpha}\Box F^{\alpha}
+ \bar{C}^{\tau}F_{\tau}^{\mu\nu}D_{\mu\nu}^{\alpha}C_{\alpha}
\end{eqnarray}

we note first of all that the gauge-fixing term is not renormalized if the
linear gauge is used (see e.g. \cite{stelle,voronovtutin}). Also the
renormalization of the ghost part of the effective action is immaterial
as long as only the one-loop expression is needed in (\ref{22-11}).

Thus to calculate $\Omega_{1/\varepsilon^2}^{[1]div(2)}$ we rewrite $L_{gf} + x A_{11}(L)$ as

\begin{eqnarray}&&
\sqrt{-g}\biggl\{(\alpha_1 + x c_2) R^2 + (\beta +x c_1) R_{\mu\nu} R^{\mu\nu}
\nonumber\\ &&
- \frac{1}{(\frac{1}{k^2} - x c_3)^{- 1}} \left(R - 2 \left[(\frac{1}{k^2} - x c_3)^{- 1}\left(\frac{\Lambda}{k^2} + \frac{x c_4 \Lambda + x c_5}{2}\right)\right]\right)\biggl\},
\end{eqnarray}

apply the one-loop result (\ref{answerr2gen}), differentiate it with respect to $x$ and set $x$ = 0.

The result of this calculation is presented in Appendix B.

Now we are in position to verify the identity (\ref{res2b}).

As follows from the result (\ref{twoloop}) on the mass-shell the left hand side of (\ref{res2b}) is

\begin{eqnarray}
&&\hspace{-1,5cm}
2\Delta\frac{d\Omega_{1/\varepsilon^2}^{[1]div(2)}}{d\Delta} =
\frac{1}{\left( 32\pi^2\varepsilon \right)^2} \int d^4 x\ \sqrt{-g}
\left( A \Lambda + B \right),
\end{eqnarray}

\begin{eqnarray}
A = \frac{1}{\beta^2 k^2}\left\{ 2 \Delta^2 w^2 + \Delta w ( - 20 v \alpha^{-1} + 20 v + 3 w \alpha^{-1} + 40/3
\alpha - 1/3 \alpha^{-1} - 10) \right\},
\nonumber
\end{eqnarray}

\begin{eqnarray}
B = - \frac{\Delta w}{\beta^3 k^4} ( 1/4 \alpha^{-2} + 5), ~w \equiv u + 7 v,
\nonumber
\end{eqnarray}

while (\ref{answerr2gen}) gives for the right hand side

\begin{eqnarray}
- a^{(1)} g_{\mu\nu}\frac{\delta\Omega^{div(1)}}{\delta g_{\mu\nu}} =
\frac{a^{(1)}}{32\pi^2\varepsilon} \int d^4 x\ \sqrt{-g}\left( \tilde{A}\Lambda + \tilde{B} \right),
\end{eqnarray}

$$\tilde{A} = \frac{1}{\beta k^2}\left\{2 \Delta w - 20 v \alpha^{-1} + 20 v
+ 3 w \alpha^{-1} + 40/3 \alpha - 1/3 \alpha^{-1} - 10\right\},$$

$$\tilde{B} = - \frac{1}{\beta^2 k^4}\left\{ 1/4 \alpha^{-2} + 5 \right\}.$$

We see that the eq.(\ref{res2b}) is really satisfied and the coefficient
$a^{(1)}$ turns out to be equal to $\frac{\Delta w}{32 \pi^2 \varepsilon\beta}.$
Note that $\frac{d\Omega^{[1]div(2)}}{d\Delta}$ is not zero even if the
unweighted gauge condition $\Delta \to 0$ is used.
\vspace{1cm}

{\bf Conclusion}

\vspace{1cm}

We have shown in this paper that generally the divergent parts of the effective action of
$R^2$-gravity depend on the gauge and parametrization nontrivially $-$ this dependence can not
be presented as proportional to the equations of motion. The renormalization
procedure in the presence of the new anticommuting source Y turned out to be
more complicated then the usual one: the renormalization equation
corresponding to the modified generating functional can not be obtained
by naive extracting of divergent terms in Slavnov identities.
We have considered the renormalization of modified Green functions
at one- and two-loop levels and obtained renormalization
equations corresponding to the insertion of Y-source (eq.(\ref{res1},\ref{res2})).
Also explicit calculation of the one- and
two-loop divergent parts has been carried out confirming our results and
demonstrating that the nontrivial gauge dependence of the divergent parts of the effective action actually exists in
arbitrary (Lorentz) gauge and arbitrary parametrizations except those
satisfying $w = 0$\footnote{Construction of the parametrization satisfying $w = 0$ see in \cite{kps1,kps2}.}.

We emphasize that this nontrivial dependence is due to the presence
of the Einstein term in the Lagrangian. Had we considered a theory
with the Lagrangian containing the higher derivative terms only
we wouldn't have had such a dependence.

Our conclusion does not contradict the equivalence theorem \cite{ktut} in view of
the general results of \cite{lavtut,tut}. Their validity in the present case
is verified in appendix B. However, these results do not allow to
say that the renormalization of the coupling constants is independent from
the renormalization of fields (as in the case of two-dimensional chiral theories \cite{tut} for example),
because renormalization of the Newtonian gravitational
constant $k$ can not be separated from renormalization of the gravitational
field: one can always perform additional redefinitions of the constant $k$ and the metric field
which compensate each other. This is a consequence of the fact that $k$ is an "inessential" coupling constant.

Finally, we note that our results are in agreement with the general statements
of \cite{hooft1}.

\vspace{0,5cm}

{\bf Acknowledgements}

We would like to thank our colleagues at the department of Theoretical Physics, Moscow State University, for many useful discussions.

We are also grateful to V.~V.~Asadov for substantial financial support of our research.

\vspace{0,5cm}

{\bf Appendix A}

\vspace{0,5cm}

In this appendix we present calculation of the one-loop divergent part
of the effective action in the flat space-time.

According to algorithm derived in \cite{algornonmin} we should first
calculate the part of $\delta^2 S_{gf}$ with four derivatives

\begin{eqnarray}&&
\left.\delta^2 S_{gf}\right|_{4} =  h\Box^2 h \left(\alpha_1 + \frac{\beta}{4} + \frac{1}{2\Delta} \right) + \frac{\beta}{4} h_{\mu\nu}\Box h^{\mu\nu}  +
A_{\nu}\Box A^{\nu}\left(\frac{\beta}{2} - \frac{1}{2\Delta} \right)
\nonumber\\ &&
+ (\nabla A)\Box h \left\{\frac{1}{\Delta}
- 2 \left(\alpha_1 + \frac{\beta}{4}\right)\right\} + (\nabla A)^2 \left(\alpha_1 + \frac{\beta}{2}\right), ~A_{\nu} \equiv \nabla^{\mu} h_{\mu\nu}.
\end{eqnarray}

Then we substitute $\nabla_{\mu} \to n_{\mu},$ $n_{\mu}$ being a vector with $n^{2}_{\mu} = 1$,
and calculate the "propagator" $(Kn^{-1})^{\alpha\beta,\gamma\delta}$ which is the inverse of the operator $(Kn)_{\mu\nu,\alpha\beta} = \frac{\delta^2 S_{gf}}{\delta h_{\mu\nu}\delta h_{\alpha\beta}}:$
$$(Kn)_{\mu\nu,\alpha\beta} (Kn^{-1})^{\alpha\beta,\gamma\delta} = \delta_{\mu\nu}^{\gamma\delta},$$

$(Kn^{-1})^{\alpha\beta,\gamma\delta} = 1/2 (g^{\alpha\gamma} g^{\beta\delta} + g^{\alpha\delta} g^{\beta\gamma}) +
A_1 g^{\alpha\beta} g^{\gamma\delta} + B_1 (g^{\alpha\gamma} n^{\beta} n^{\delta} + g^{\alpha\delta} n^{\beta} n^{\gamma} + g^{\beta\gamma} n^{\alpha} n^{\delta} + g^{\beta\delta} n^{\alpha} n^{\gamma}) +
C_1 (g^{\alpha\beta} n^{\gamma} n^{\delta} + g^{\gamma\delta} n^{\alpha} n^{\beta}) +
D_1 n^{\alpha} n^{\beta} n^{\gamma} n^{\delta},$
where
$$A_1 = - (A + 4 A B + A D - C^2)/Z,$$
$$B_1 = - B/(1 + 2 B),$$
$$C_1 = (4 A B + A D - C - C^2)/Z,$$
$$D_1 = - (16 A B + 4 A D + D + 4 B - 4 C^2)/Z + 4 B/(1 + 2 B),$$
$$Z = 1 + 4 A + 4 B + 2 C + D + 3 (4 A B + A D - C^2),$$
coefficients $A, B, C, D$ being defined from $L_{gf}$:
$$A = 1 + \frac{4\alpha_1}{\beta} + \frac{2}{\beta\Delta},$$
$$B = \frac{1}{2}\left( \frac{1}{\beta\Delta} - 1 \right),$$
$$C = - 1 - \frac{4\alpha_1}{\beta} + \frac{2}{\beta\Delta},$$
$$D = 2 + \frac{4\alpha_1}{\beta}.$$

We have multiplied the initial Lagrangian by $\frac{4}{\beta}$ for convenience.

Second, we calculate the part $W$ of $\delta^2 S_{gf}$ containing two
derivatives substituting $\nabla_{\mu} \to n_{\mu}$ again

\begin{eqnarray}&&
(Wn)_{\mu\nu,\alpha\beta} = \frac{1}{k^2\beta}\left\{g_{\mu\nu} g_{\alpha\beta} -
(g_{\mu\nu} n_{\alpha} n_{\beta} + g_{\alpha\beta} n_{\mu} n_{\nu}) -
\right.
\nonumber\\&&
\left.
g_{\mu\alpha} g_{\nu\beta} + (g_{\mu\beta} n_{\nu} n_{\alpha} + g_{\nu\alpha} n_{\mu} n_{\beta}) \right\}
\end{eqnarray}

and the part $M$ without derivatives

\vspace{0,5cm}

$M_{\mu\nu,\alpha\beta} = \frac{4 \Lambda u}{\beta k^2}g_{\mu\nu} g_{\alpha\beta} +
\frac{4\Lambda v}{\beta k^2}g_{\mu\alpha} g_{\nu\beta},$

\vspace{0,5cm}

where $u = a + 4 c + \frac{1}{4}, v = b + 4 d - \frac{1}{2}.$

The one-loop divergent part of the effective action has the form\footnote{Since the space-time is flat the contributions of the
Faddeev-Popov ghosts and of the "third" ghost are equal to zero.}

$$\Omega^{div(1)} =
\frac{1}{32\pi^2\varepsilon}tr \int d^4 x\ \sqrt{-g}\left( \frac{1}{2}(Kn^{-1})(Wn)(Kn^{-1})(Wn) - (Kn^{-1})(M)\right),$$

where the matrix product of $(Kn^{-1})_{\mu\nu,\alpha\beta}, (Wn)_{\mu\nu,\alpha\beta}, M_{\mu\nu,\alpha\beta}$
is supposed.

A simple calculation gives

\begin{eqnarray}
\Omega^{div(1)}_{flat} =
\frac{1}{32\pi^2\varepsilon} \int d^4 x\ \sqrt{-g}\left( c_4 \Lambda + c_5 \right),
\end{eqnarray}

where

\begin{eqnarray}
c_4 &=&
- \frac{1}{\beta k^2}\left[ u \left( 2\Delta + \frac{3}{\alpha} \right) + v \left( 14 \Delta + \frac{1}{\alpha} + 20 \right) \right],
\nonumber
\end{eqnarray}

$$c_5 = \frac{1}{\beta^2 k^4}\left(\frac{5}{2} + \frac{1}{8\alpha^2}\right).$$

\pagebreak
\vspace{1cm}

{\bf Appendix B}

\vspace{0,5cm}

In this appendix the result of calculation of the two-loop divergent
as $\frac{1}{\varepsilon^2}$ part of the effective action is presented.
Also, validity of the general statements of \cite{lavtut,tut} is verified.

Following the algorithm derived in sec.4b we obtain

\begin{eqnarray}\label{twoloop}
&&\hspace{-1,5cm}
\Omega_{1/\varepsilon^2}^{[1]div(2)} =
\frac{1}{2\left( 32\pi^2\varepsilon \right)^2} \int d^4 x\ \sqrt{-g}
\left( c_{22} R^2 + c_{32} R + c_{42} \Lambda + c_{52} \right),
\end{eqnarray}

\begin{eqnarray}
c_{22} = \frac{1}{\beta}(200/27 \alpha^3 - 416/9 \alpha^2 + 1697/54 \alpha - 25/36),
\nonumber
\end{eqnarray}
\vspace{-0,5cm}
\begin{eqnarray}&&
c_{32} = \frac{1}{\beta^2 k^2}\left\{ \Delta^2 ( - 1/4 u^2 - 7/2 u v - 49/4 v^2)
\right.
\nonumber\\&&
\left.
+ \Delta ( - 3/4 u^2\alpha^{-1} - 11/2 u v \alpha^{-1} - 5 u v + 10/3 u \alpha
\right.
\nonumber\\&&
\left.
- 1/12 u \alpha^{-1} - 183/20 u - 7/4 v^2 \alpha^{-1} - 35 v^2 + 70/3 v \alpha
\right.
\nonumber\\&&
\left.
- 7/12 v \alpha^{-1} - 1281/20 v) - 9/16 u^2 \alpha^{-2} - 15/2 u v \alpha^{-1}
\right.
\nonumber\\&&
\left.
- 3/8 u v \alpha^{-2} - 7/16 u \alpha^{-2} + 5/2 u - 5/2 v^2 \alpha^{-1}
\right.
\nonumber\\&&
\left.
- 1/16 v^2 \alpha^{-2} - 25 v^2 + 100/3 v \alpha - 5/6 v \alpha^{-1} - 7/48 v \alpha^{-2}
\right.
\nonumber\\&&
\left.
- 272/3 v - 200/9 \alpha^2 + 122 \alpha - 1/24 \alpha^{-2} - 731/18 \right\},
\nonumber
\end{eqnarray}

\vspace{-0,5cm}

\begin{eqnarray}&&
c_{42} = \frac{1}{\beta^2 k^2}\left\{ 2 \Delta^2 (u^2 + 14 u v + 49 v^2)
\right.
\nonumber\\&&
\left.
+ \Delta (6 u^2 \alpha^{-1} + 44 u v \alpha^{-1} + 40 u v + 133/5 u + 14 v^2 \alpha^{-1}
\right.
\nonumber\\&&
\left.
+ 280 v^2 + 931/5 v) + 9/2 u^2 \alpha^{-2} + 60 u v \alpha^{-1} + 3 u v \alpha^{-2}
\right.
\nonumber\\&&
\left.
- 15 u \alpha^{-1} + 5/4 u \alpha^{-2} + 10 u + 20 v^2 \alpha^{-1} + 1/2 v^2 \alpha^{-2}
\right.
\nonumber\\&&
\left.
+ 200 v^2 - 5 v \alpha^{-1} + 5/12 v \alpha^{-2} + 808/3 v \right\},
\nonumber
\end{eqnarray}

\vspace{-0,5cm}

\begin{eqnarray}&&
c_{52} = \frac{1}{\beta^3 k^4}\left\{ \Delta ( - 1/4 u \alpha^{-2} - 5 u - 7/4 v \alpha^{-2} - 35 v)
\right.
\nonumber\\&&
\left.
- 15/2 u \alpha^{-1} - 3/8 u \alpha^{-3} - 5/2 v \alpha^{-1} - 5/2 v \alpha^{-2} - 1/8 v \alpha^{-3}
\right.
\nonumber\\&&
\left.
- 50 v + 50/3 \alpha - 5/12 \alpha^{-1} + 5/8 \alpha^{-2} - 1/8 \alpha^{-3} - 79 \right\}.
\nonumber
\end{eqnarray}

To show the gauge and parametrization dependence can be absorbed by a field
renormalization we first remove one-loop divergencies (\ref{answerr2gen})
by the following redefinition of charges and fields

$$ g_{\mu\nu} \to g_{\mu\nu} (1 + \delta_1 Z),$$
$$\lambda \to \lambda (1 + \delta_1 \lambda),
~\frac{1}{k^2} \to \frac{1}{k^2}(1 + \delta_1 \frac{1}{k^2}),$$
$$\alpha_1 \to \alpha_1 (1 + \delta_1 \alpha_1),
~\beta \to \beta (1 + \delta_1 \beta),$$

where

$$\delta_1\frac{1}{k^2} + \delta_1 Z = \frac{c_3}{32\pi^2\varepsilon},$$
$$\delta_1\lambda = - \frac{2 c_3^i + c_5/2\lambda}{32\pi^2\varepsilon},$$
$$\delta_1\alpha_1 = \frac{- c_2}{32\pi^2\varepsilon\alpha_1},
~\delta_1\beta = \frac{- c_1}{32\pi^2\varepsilon\beta},$$

and we have introduced a notation $c_3^i$ for the gauge and parametrization
independent part of the coefficient $c_3$
$$ c_3 \equiv c_3^i - \frac{c_4}{4}.$$

As seen from the above equations renormalizations of the gravitational
constant and of the metric field can not be separated from each other.
This property is inherent to any metrical theory of gravity with the
Lagrangian containing terms linear in curvature and holds
at any order of perturbation theory.

To make the theory finite at the two-loop level we should take into account
counterterms which arise in the second order from the one-loop redefinitions
of the charges and fields which were made above.
Correspondingly, we extract these counterterms from the two-loop order
result (\ref{twoloop}) rewriting coefficients $c_{32}, c_{42}$ and $c_{52}$
as\footnote{The one-loop redefinitions do not affect the coefficient $c_{22}$.}

\begin{eqnarray}&&
c_{32} = \frac{k^2}{2}c_3^i c_4 +
\frac{1}{\beta^2 k^2}\{ \Delta^2 ( - 1/4 u^2 - 7/2 u v - 49/4 v^2)
+ \Delta ( - 3/4 u^2 \alpha^{-1}
\nonumber\\&&
- 11/2 u v \alpha^{-1}
- 5 u v - 133/20 u - 7/4 v^2 \alpha^{-1} - 35 v^2 - 931/20 v) - 9/16 u^2 \alpha^{-2}
\nonumber\\&&
- 15/2 u v \alpha^{-1} - 3/8 u v \alpha^{-2} + 15/4 u \alpha^{-1} - 5/16 u \alpha^{-2} - 5/2 u - 5/2 v
^2 \alpha^{-1}
\nonumber\\&&
- 1/16 v^2 \alpha^{-2} - 25 v^2 + 5/4 v \alpha^{-1} - 5/48 v \alpha^{-2} - 202/3 v - 200/9 \alpha^2
\nonumber\\&&
+ 122 \alpha - 1/24 \alpha^{-2} - 731/18\},
\nonumber
\end{eqnarray}

\vspace{-0,5cm}

\begin{eqnarray}&&
c_{42} = k^2 \{c_4^2/4 - 12 c_{3i}^2 \} +
\frac{1}{\beta^2 k^2}\{ \Delta^2 (u^2 + 14 u v + 49 v^2) + \Delta (3 u^2 \alpha^{-1} + 22 u v \alpha^
{-1}
\nonumber\\&&
+ 20 u v + 133/5 u + 7 v^2 \alpha^{-1} + 140 v^2 + 931/5 v) +
 9/4 u^2 \alpha^{-2} + 30 u v \alpha^{-1} + 3/2 u v \alpha^{-2}
\nonumber\\&&
- 15 u \alpha^{-1} + 5/4 u \alpha^{-2} + 10 u + 10 v^2 \alpha^{-1} + 1/4 v^2 \alpha^{
-2} + 100 v^2 - 5 v \alpha^{-1} + 5/12 v \alpha^{-2}
\nonumber\\&&
+ 808/3 v + 400/3\alpha^2 - 200 \alpha + 5 \alpha^{-1} + 1/12 \alpha^{-2} + 205/3 \},
\nonumber
\end{eqnarray}

\vspace{-0,5cm}

\begin{eqnarray}
c_{52} = k^2 c_5 (c_4 - 4 c_3^i)
- \frac{1}{\beta^3 k^4}\{ 50/3 \alpha + 5/4 \alpha^{-1} - 15/8 \alpha^{-2} + 1/12 \alpha^{-3} + 54 \}.
\nonumber
\end{eqnarray}

Now it is easy to verify that

\begin{eqnarray}&&
4 c_{32} + c_{42} - k^2 (2 c_{3i} c_4 + c_4^2/4 - 12 c_{3i}^2) =
\nonumber\\&&
\frac{1}{\beta^2 k^2}\{ 400/9 \alpha^2 + 288 \alpha + 5 \alpha^{-1} - 1/12 \alpha^{-2} - 847/9 \},
\nonumber
\end{eqnarray}

which means that after subtraction of the counterterms corresponding to
the one-loop renormalization of charges and fields is made the two-loop
divergent part of the effective action becomes gauge and parametrization
independent on-shell. Therefore the gauge and parametrization
dependence can be absorbed by a field renormalization or by renormalization
of the Newtonian constant.



\begin{thebibliography}{99}

\bibitem{wein}
~S.~Weinberg, in: "General Relativity", eds. ~S.~W.~Hawking and ~W.~Israel
(Cambridge Univ. Press, Cambridge, 1979).

\bibitem{lavtut}
~P.~M.~Lavrov,~I.~V.~Tyutin,~B.~L.~Voronov,
Sov. Journ. of Nucl. Phys., {\bf v.36}, n.2, (1982).

\bibitem{batvil1}
~I.~A.~Batalin, ~G.~A.~Vilkovisky, Phys. Lett., {\bf 69B}, 309, (1977).

\bibitem{batvil2}
~I.~A.~Batalin, ~G.~A.~Vilkovisky, Phys. Lett., {\bf 102B}, 27, (1981).

\bibitem{batvil3}
~I.~A.~Batalin, ~G.~A.~Vilkovisky, Nucl. Phys., {\bf 234}, 106, (1984).


\bibitem{kluberg1}
~H.~Kluberg-Stern and ~J.~B.~Zuber, Phys. Rev., D {\bf 12}, (1975), 467.

\bibitem{kluberg2}
~H.~Kluberg-Stern and ~J.~B.~Zuber, Phys. Rev., D {\bf 12}, (1975), 3159.

\bibitem{stelle}
~K.~S.~Stelle, Phys. Rev., D {\bf 16}, n. 4, (1977), 953.

\bibitem{zinnjustin}
~J.~Zinn-Justin, in: "Trends in Elementary Particle Theory", v.37, eds.
Rollnik and Dietz, Springer Verlag, 1975.

\bibitem{voronovtutin}
~I.~V.~Tyutin,~B.~L.~Voronov,
Sov. Journ. of Nucl. Phys.,{\bf v.39}, n.4, 998, (1984).

\bibitem{dewitt}
~B.~S.~DeWitt, Phys. Rev., {\bf 162}, (1967), 1195.

\bibitem{hooft}
~G.'t Hooft, Nucl. Phys., {\bf  B62}, 444, (1973).

\bibitem{hooftvelt}
~G.'t Hooft and ~M.~Veltman, Ann. Inst. ~H.~Poincare, {\bf 20}, (1974), 69.

\bibitem{avrbar}
~I.~G.~Avramidy and ~A.~O.~Barvinsky, Phys. Lett., v.{\bf 159B}, nn. 4, 5, 6., 269, (1985).

\bibitem{blox}
~D.~I.~Blohintzev, ~A.~V.~Efremov, ~D.~V.~Shirkov, Izv. Vuzov, ser. Fizika, {\bf 12}, 23, (1974).

\bibitem{kaz}
~D.~I.~Kazakov, Teor. Mat. Fiz, {\bf v.75}, n.1, 157, (1988).

\bibitem{algornonmin}
~P.~I.~Pronin, ~K.~V.~Stepanyantz, Nucl. Phys., {\bf B 485}, 517, (1997).

\bibitem{kps1}
~K.~A.~Kazakov, ~P.~I.~Pronin, ~K.~V.~Stepanyantz, Gravitation and Cosmology, n.1, 17, (1998).

\bibitem{kps2}
~M.~Yu.~Kalmykov, ~K.~A.~Kazakov, ~P.~I.~Pronin,
in: Proceedings of the "First Open Scientific Conference of Young Specialists",
JINR, Dubna, p.65, (1997).

\bibitem{ktut}
~R.~Kallosh, ~I.~V.~Tyutin,
Sov. Journ. of Nucl. Phys., {\bf v.17}, 190, (1973).

\bibitem{tut}
~I.~V.~Tyutin,
Sov. Journ. of Nucl. Phys., {\bf v.35}, n.1, (1982).

\bibitem{hooft1}
t'Hooft G., Quantum gravity. - Lect. Notes. Phys., {\bf v.37}, 92, (1975).

\end{thebibliography}
\end{document}